# Efficient All-Optical Helicity Dependent Switching of Spins in a Pt/Co/Pt Film by a Dual-Pulse Excitation

**Kihiro T. Yamada[1,2*], Alexey V. Kimel[1*], Kiran Horabail Prabhakara[1], Sergiu Ruta[3], Tian Li[4], Fuyuki Ando[4], Sergey Semin[1], Teruo Ono[4,5], Andrei Kirilyuk[1,6], Theo Rasing[1]**

[1]Radboud University, Institute for Molecules and Materials, Nijmegen 6525 AJ, The Netherlands

[2]Department of Physics, Tokyo Institute of Technology, Tokyo 152-8551, Japan

[3]Department of Physics, University of York, Heslington, York YO10 5DD, United Kingdom

[4]Institute for Chemical Research, Kyoto University, Uji, Kyoto 611-0011, Japan

[5]Center for Spintronics Research Network (CSRN), Graduate School of Engineering Science, Osaka University, Toyonaka, Osaka 560-8531, Japan

[6]FELIX Laboratory, Radboud University, Toernooiveld 7c, Nijmegen 6525 ED, The Netherlands

*** Correspondence:**
Kihiro T. Yamada
yamada@phys.titech.ac.jp
Alexey V. Kimel
aleksei.kimel@ru.nl

**Abstract**

All-optical helicity dependent switching (AO-HDS), deterministic control of magnetization by circularly polarized laser pulses, allows to efficiently manipulate spins without the need of a magnetic field. However, AO-HDS in ferromagnetic metals so far requires many laser pulses for fully switching their magnetic states. Using a combination of a short, 90-fs linearly polarized pulse and a subsequent longer, 3-ps circularly polarized pulse, we demonstrate that the number of pulses for full magnetization reversal can be reduced to 4 pulse pairs in a single stack of Pt/Co/Pt. The obtained results suggest that the dual-pulse approach is a potential route towards realizing efficient AO-HDS in ferromagnetic metals.

## 1 Introduction

The explosive growth of big data and artificial intelligence demands faster and more energy-efficient ways to manipulate and store data (Manipatruni et al., 2018; Beyond CMOS, 2018; B. Dieny et al., 2020). One approach is to use current-induced torque (A. Brataas et al., 2012; B. Dieny et al., 2020) to switch magnetization, as in spin-transfer torque magnetic random access memories (Apalkov et al., 2016). Another potential route is to use an ultrashort laser pulse (Kirilyuk et al., 2010; Kimel and Li., 2019). Merging the ultrafast and energy-efficient features of photonics with the nonvolatility of magnets may lead to innovations in information technologies. Even though the leading part of all-optical magnetization switching has been ferrimagnets, represented by GdFeCo (Stanciu et al., 2007;

Vahaplar et al., 2009; Radu et al., 2011; Ostler et al., 2012; Khorsand et al., 2012; Mangin et al., 2014), cutting-edge magnetic media consist of ferromagnets (Apalkov et al., 2016; Kryder et al., 2008). This is the reason why all-optical switching in ferromagnets attracts great attention. Ferromagnets, such as Pt/Co/Pt multilayers (Lambert et al., 2014; Hadri et al., 2016; Hadri et al., 2016; Medapalli et al., 2017; Quessab et al., 2018; Parlak et al., 2018; Kichin et al., 2019; Chakravarty et al., 2019; Cheng et al., 2020) and granular FePt alloys (Lambert et al., 2014; Takahashi et al., 2016; John et al., 2017), have recently shown all-optical helicity-dependent switching (AO-HDS), magnetization switching depending on the circular polarization of the optical excitation pulses. However, AO-HDS generally requires at least tens and usually hundreds of laser pulses for full switching (Kichin et al., 2018).

AO-HDS is understood to proceed via two stages: helicity-independent nucleation of switched domains and helicity-dependent deterministic domain wall (DW) motion (Hadri et al., 2016; Medapalli et al., 2017). This two-step mechanism especially motivated us to try a dual-pulse excitation in which each pulse of a pair of pulses is optimized for either the helicity independent demagnetization and nucleation of switched domains or the helicity dependent DW motion. Hence, the goal of our study was to find the optimal combination of pulses that allows switching the magnetization of ferromagnetic materials in a helicity dependent and deterministic way. The studied Pt/Co/Pt trilayer with perpendicular magnetic anisotropy is excited by pairs of laser pulses. The pair consisted of a femtosecond (fs) linearly polarized ($\pi$) laser pulse, meant to demagnetize the sample, and a picosecond (ps) circularly polarized ($\sigma$) laser pulse coming after a time separation ($\Delta t$), meant to steer the magnetization to its final state in a helicity dependent way, but without causing further demagnetization. We found that using only four of such dual pulse pairs are sufficient to deterministically reverse the magnetization, offering great perspectives for AO-HDS of ferromagnetic media.

## 2 Materials and Methods

### 2.1 Sample preparations

We prepared a Ta (4 nm)/Pt (3.0 nm)/Co (0.6 nm)/Pt (3.0 nm)/MgO (2.0 nm)/Ta (1.0 nm) multilayer on a synthetic quartz glass substrate by magnetron sputtering. Pt/Co/Pt systems are relevant for testing the dual-pulse approach because they are typical candidates for spintronic devices (A. Brataas et al., 2012) as well as for all-optical helicity dependent switching (Lambert et al., 2014; Hadri et al., 2016; Hadri et al., 2016; Medapalli et al., 2017; Quessab et al., 2018; Parlak et al., 2018; Kichin et al., 2019; Chakravarty et al., 2019; Cheng et al., 2020). DC and RF sources were used for depositing Ta, Pt, and Co, and MgO, respectively. The MgO/Ta capping layer prevents the magnetic layer from oxidization. The multilayer exhibits a perpendicular easy axis of magnetization.

### 2.2 Magneto-optical imaging

For optical excitation, we used a Ti: sapphire amplified laser system (Solstice Ace, Spectra-Physics) of which the central wavelength and repetition rate were 800 nm and 1 kHz, respectively. We used two compressors of the amplifier system to tune the pulse widths of the two pump laser beams. The laser amplifier was used in external trigger mode. Using voltage pulses from a delay generator (DG645, Stanford Research), we controlled the number of pulses for pumping. The durations of the short and long pump pulses were about 90 fs and 3.0 ps, respectively. The time separation, $\Delta t$, was adjusted by a delay line on the pathway of the long pulse. The laser pulses with Gaussian intensity distribution were incident at an angle of 15 deg from the sample normal. The focused beam sizes ($1/e^2$ radius) of the pump pulses were calculated with the Liu method (Khorsand et al., 2012; Liu, 1982). The laser fluence was calculated using the $1/e^2$ radius, the repetition rate (1 kHz) and the average power measured with a power meter. The pump intensities were controlled using a combination of a half-wave plate and a Gran-Taylor prism. To obtain a circular polarized beam, we used a quarter-wave plate (AQWP05M-600, Thorlabs Inc.). We used a white light source to visualize the magnetic domains in transmission via the magneto-optical Faraday effect. The white light source, which was linearly polarized by a sheet polarizer, was collimated and incident on the sample surface using a combination of lenses. We collected the transmitted light from the sample by an objective with a magnification of 20×. The polarization axis of an analyzer was nearly orthogonal to the polarization of the incident probe light. Since the magnetized sample rotates the polarization of the incident light, the magnetic domains can be visualized by a charge-coupled device (CCD) camera placed behind the analyzer.

## 3 Results

We first measured the switching by only circularly polarized pulses in transmission. **Figure 1A** shows magneto-optical images taken after the Pt/Co/Pt film was excited with a sequence of 3.0-ps right($\sigma^+$)- or left($\sigma^-$)-handed circularly polarized light pulses. These images were taken long after the excitation when the magnetization had already reached a stable state. The time interval was more than 2 seconds. Here, we used ps σ pulses because ps σ pulses are more favorable for the AO-HDS than fs σ pulses (see Medapalli et al. 2017). To fully reverse the magnetization in the area with a diameter of 15-μm, the experiments required 100-150 pulses.

We next performed similar experiments with pairs of pulses. **Figure 1B** shows the results, in which the same Pt/Co/Pt stack was excited with one 90-fs π and one 3.0–ps σ pulses, separated by $\Delta t$ = 5.0 ps. **Figure 1B** shows that 4 pairs of these dual pulses can completely switch the magnetizations of the same 15-μm area in a deterministic way. The switched area can be switched back by switching the helicity of the longer pulse. We also studied the impact of the pulse width of the σ pulse on the dual-pulse AO-HDS and found that longer σ pulses are better not only for the multi-pulse AO-HDS but also for the dual-pulse AO-HDS. We again note that in the case of excitation without π pulses, i.e., with σ pulses only, a similar result would require more than 100 excitation events (**Figure 1A**). The total laser fluence for full magnetization switching with the dual-pulse method, $4 \times (F_\pi + F_\sigma) \sim 17$ mJ/cm$^2$, is much smaller than $100 \times F_\sigma \sim 342$ mJ/cm$^2$ used for the multi single-pulse approach. The dual-pulse approach not only minimizes the number of pulses but also dramatically reduces the net laser fluence required for full magnetization reversal.

The pulse duration is critical for the multi-pulse AO-HDS. We checked that the pulse width of the long pulse must be pico-second wide. **Figure 2** show the snapshots when using a σ pulse with a duration $\tau_\sigma$ of 0.5 ps, 1.0 ps, 2.0 ps, and 3.0 ps for $\Delta t$ = 5.0 ps. Full magnetization reversal is enabled with increasing $\tau_\sigma$. The duration of the σ pulse must be more than 1 ps to obtain a clear helicity dependent effect. This may be because a short pulse makes the spin temperature reach the Curie temperature.

We next checked if a circular polarization of the fs pulse has an impact on the dual-pulse AO-HDS. Snapshots in **Figure 3** show the dual-pulse AO-HDS by a combination of 90-fs and 3.0-ps σ pulses with $\Delta t$ = 5.0 ps. One cannot find any meaningful impact of the helicity of the fs pulses on the final magnetization state. The results show that the first process of the dual pulse AO-HDS is helicity independent, like multi-pulse AO-HDS. Hundreds of fs after the act of the short pulse, the spin temperature reaches the Curie temperature and magnetic textures should be thermally nucleated. Therefore, the helicity of the fs pulse does not largely boost or suppress the creation of magnetic textures, although the absorption of the fs σ pulse slightly changes according to the initial magnetization orientation and the light helicity due to magnetic circular dichroism (MCD) (Tsema, et al., 2016).

## 4 **Discussion**

The results above clearly demonstrate the advantage of the dual-pulse AO-HDS and trigger the question of the mechanism behind it. It is known that the inverse Faraday effect (Kimel et al., 2005) is able to switch the magnetization. Atomistic spin simulations show however that the effective field of the inverse Faraday effect must be tens of Tesla to deterministically reverse the spins of FePt (Ellis, et al., 2016). To realize such large field amplitude in our sample, we must use a fluence much large than the damage threshold for our sample. Previous studies indicated that the inverse Faraday effect can induce coherent magnetization dynamics, but the magnetization precession angles are small with similar fluences (Choi et al., 2017). If the inverse Faraday of our sample is strong, it is strange that the helicity of the fs pulse has little impact on the dual-pulse AO-HDS. Therefore, the inverse Faraday effect is unlikely to drive the dual-pulse AO-HDS.

Another probable mechanism is helicity-dependent laser absorption via MCD (Gorchon et al, 2016). MCD can produce a heat gradient across a DW, and subsequently push a DW from the hotter (lower DW energy) to the colder (higher DW energy) area (Hadri et al., 2016; Hadri et al., 2016; Medapalli et al., 2017; Quessab et al., 2018, Parlak et al., 2018), as the up (down)-magnetized domain absorbs the left (right) circularly polarized light more than the down (up)- magnetized domain. During the cooling process, the DW will not move back to the initial position due to pinning. When a circularly polarized pulse with $F_{\sigma} = 1.94$ mJ/cm$^2$ illuminates the Pt/Co/Pt stack, the heat gradient across a DW can be estimated to be $\nabla T = \Delta A_{\mathrm{MCD}} F_{\sigma} / Clt$ ~ 2.7 K/nm. Here, we used a typical DW width $l = 5.5$ nm (Metaxas et al., 2007), an MCD coefficient $\Delta A_{\mathrm{MCD}} = 0.015$ (Medapalli et al., 2017), a weighted heat capacity $C = 2.93$ MJ/Km$^3$(Rumble, 2017), and the total thickness of the Pt/Co/Pt multilayer $t =$ 6.6 nm. The estimated heat gradient is five orders of magnitude larger than one to move a DW of a magnetic garnet (Jiang et al., 2013). Heat gradients are not formed unless magnetic textures are present. This agrees with the fact that magnetic textures are necessary for both the multi-pulse and dual-pulse AO-HDS. Since the inverse Faraday effect and MCD effect show distinctive wavelength dependences, a wavelength dependence study of AO-HDS in ferromagnets may be valuable to reveal the dominant mechanism, as done for GdFeCo systems (Khorsand et al., 2012).

For the switching process of the dual-pulse AO-HDS, the fs pulse will contribute to the helicity independent creation of switched magnetic textures. Simultaneously, the reduction of the magnetic properties by the fs pulse should facilitate the magnetization switching, like heat-assisted magnetic recording (Kryder et al., 2008), where the reduced coercivity field by laser heating enables the magnetization switching by a smaller magnetic field. While the demagnetization proceeds within 500 fs - 1 ps, the remagnetization takes tens of ps after the act of the fs pulse (Borchert et al., 2020). The required pulse separation between 0 and 10 ps suggests that the helicity dependent absorption of the ps pulse may affect the remagnetizing system, resulting in efficient magnetization switching. Time-resolve studies and simulations on the dual-pulse AO-HDS should give crucial insights into the details of the actual switching dynamics of this new approach.

## 5 Conclusion

In conclusion, we have demonstrated that a dual-pulse method can drastically reduce the pulse number required for AO-HDS in a Pt/Co/Pt multilayer. Using a linearly polarized femtosecond pulse for the first pump and a circularly polarized picosecond pulse for the second pump, we successfully realized a record small number of 8 pulses to fully switch a magnetic domain. The small impact of the helicity of the femtosecond pulse on the dual-pulse AO-HDS suggests that the nucleation of magnetic textures by the pulse is helicity independent. The dual-pulse AO-HDS is optimized at a pulse separation of about 5ps. This implies that the switching might be completed on an ultrashort time scale by controlling the remagnetization in a helicity dependent way. The femtosecond pulse can forcibly create magnetic textures even in granular magnetic media (Takahashi et al., 2016; John et al., 2017) and nano devices (A. Brataas et al., 2012; B. Dieny et al., 2020) where a single domain state has been stabilized at room temperature. This means that the dual-pulse AO-HDS is practical for the application of opto-magnetic writing in magnetic recording media. The dual-pulse method is not only efficient but also a potential route towards implementation of all-optical magnetization switching to practical magnetic recording media, such as hard disc drives and magnetic random access memories.

## 6 Figure captions

**FIGURE 1 |** Dual-pulse all-optical helicity dependent switching (AO-HDS). (**A**) AO-HDS by multiple right($\sigma^+$) and left ($\sigma^-$) circularly polarized pulses. Here, the fluence of the 3.0-ps σ pulse was fixed at $F_\sigma$ = 3.42 mJ/cm$^2$. (**B**) Dual-pulse AO-HDS for time separation $\Delta t$ of 5.0 ps. Here, we used the fluence of $F_\pi$ = 2.32 mJ/cm$^2$ for the 90-fs π pulse and $F_\sigma$ = 1.94 mJ/cm$^2$ for the 3.0-ps σ pulse. The number of pulse (pairs) is indicated on each magneto-optical image. The darker and brighter areas denote up($M^\uparrow$)- and down($M^\downarrow$)-magnetized states, respectively. The scale bars correspond to 20 μm. The 1/e$^2$ diameter of the σ (solid line) and π (dash line) pulses were 52.0±0.8 μm and 68.2 ±1.8 μm, respectively.

**FIGURE 2 |** The pulse width $\tau_\sigma$ dependence of the dual-pulse AO-HDS. Snapshots before and after illuminating the uniformly magnetized sample with 1-5 pairs of pulses at $\Delta t$ of 5.0 ps are displayed for $\tau_\sigma$ = 0.5 ps, 1.0 ps, 2.0 ps, and 3.0 ps. Here, we used fluences $F_\pi$ = 2.29 mJ/cm$^2$ for the π pulse. The fluences of the σ pulses were $F_\sigma$ = 1.91 mJ/cm$^2$, 1.95 mJ/cm$^2$, 1.87 mJ/cm$^2$, and 1.90 mJ/cm$^2$ for $\tau_\sigma$ = 0.5 ps, 1.0 ps, 2.0 ps, and 3.0 ps, respectively. The 1/e$^2$ diameter of the σ (solid line) and π (dash line) pulses were 52.0±0.8 μm and 68.2 ±1.8 μm, respectively. The scale bar corresponds to 20 μm.

**FIGURE 3 |** AO-HDS with a combination of 90-fs and 3.0-ps circularly polarized pulses. Snapshots before and after illuminating the uniformly magnetized sample with 1-5 pairs of pulses at $\Delta t$ of 5.0 ps are displayed. The error bars were determined by repeating the same measurements three times. Here, we used fluences $F_{\sigma 1}$ = 2.35 mJ/cm$^2$ and $F_{\sigma 2}$ = 1.32 mJ/cm$^2$ for fs and ps circularly polarized pulses, respectively. The 1/e$^2$ diameter of the ps (solid line) and fs (dash line) pulses were 84.2±1.2 μm and 82.7 ±1.0 μm, respectively. The scale bar corresponds to 20 μm.

## 7 Conflict of Interest

The authors declare that the research was conducted in the absence of any commercial or financial relationships that could be construed as a potential conflict of interest.

## 8 Author Contributions

KTY, AVK, and TR conceived the experiments. KTY, KHP, and SS designed and build the experimental set-up. KTY performed the measurements, and collected data. All the data were analyzed by KTY. with the help of AVK and TR. The sample was fabricated and provided by TL, FA, and TO. KTY, AVK, and TR wrote the manuscript. All authors discussed the results and commented on the manuscript. This project was coordinated by AVK and TR.

## 9 Funding

This work was partly supported by the European Research Council Grant Agreement No.856538 (3D-MAGiC), by the FOM programme Exciting Exchange, de Nederlandse Organisatie voor Wetenschappelijk Onderzoek (NWO), by the EU H2020 Program Grant Agreement No. 713481 (SPICE) and No. 737093 (FEMTOTERABYTE), by JSPS KAKENHI No. 15H05702, No. 20H00332, No. 20H05665, No. 17J07326, and No. 18J22219, and by the Collaborative Research Program of the Institute for Chemical Research, Kyoto University.

## 10 Acknowledgments

We acknowledge T. Toonen and C. Berkhout for the continuous technical supports, J. Mentink and T. Satoh for fruitful discussions, and T. Taniguchi for valuable comments on sample fabrication.

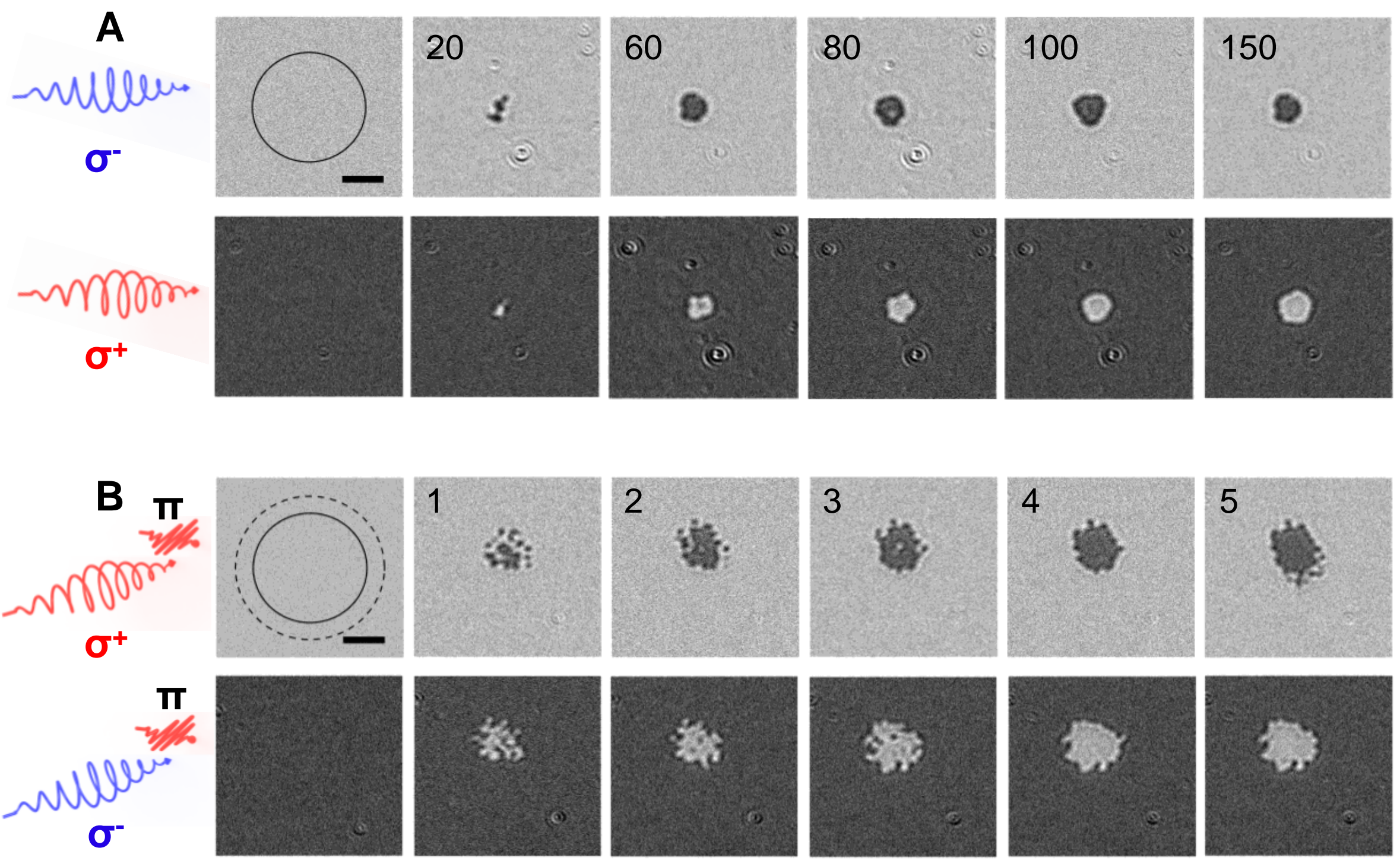

Figure 1 Yamada et al.

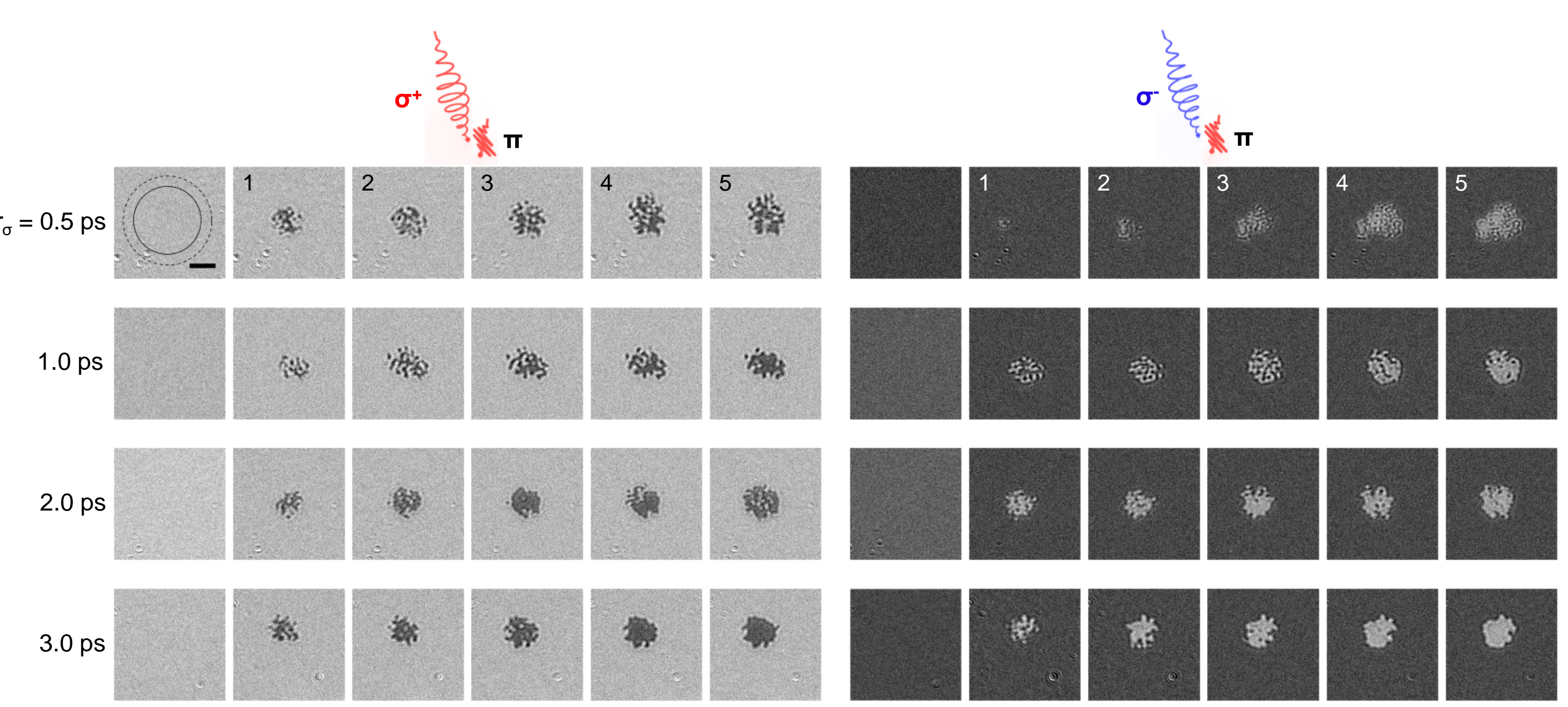

Figure 2 Yamada et al.

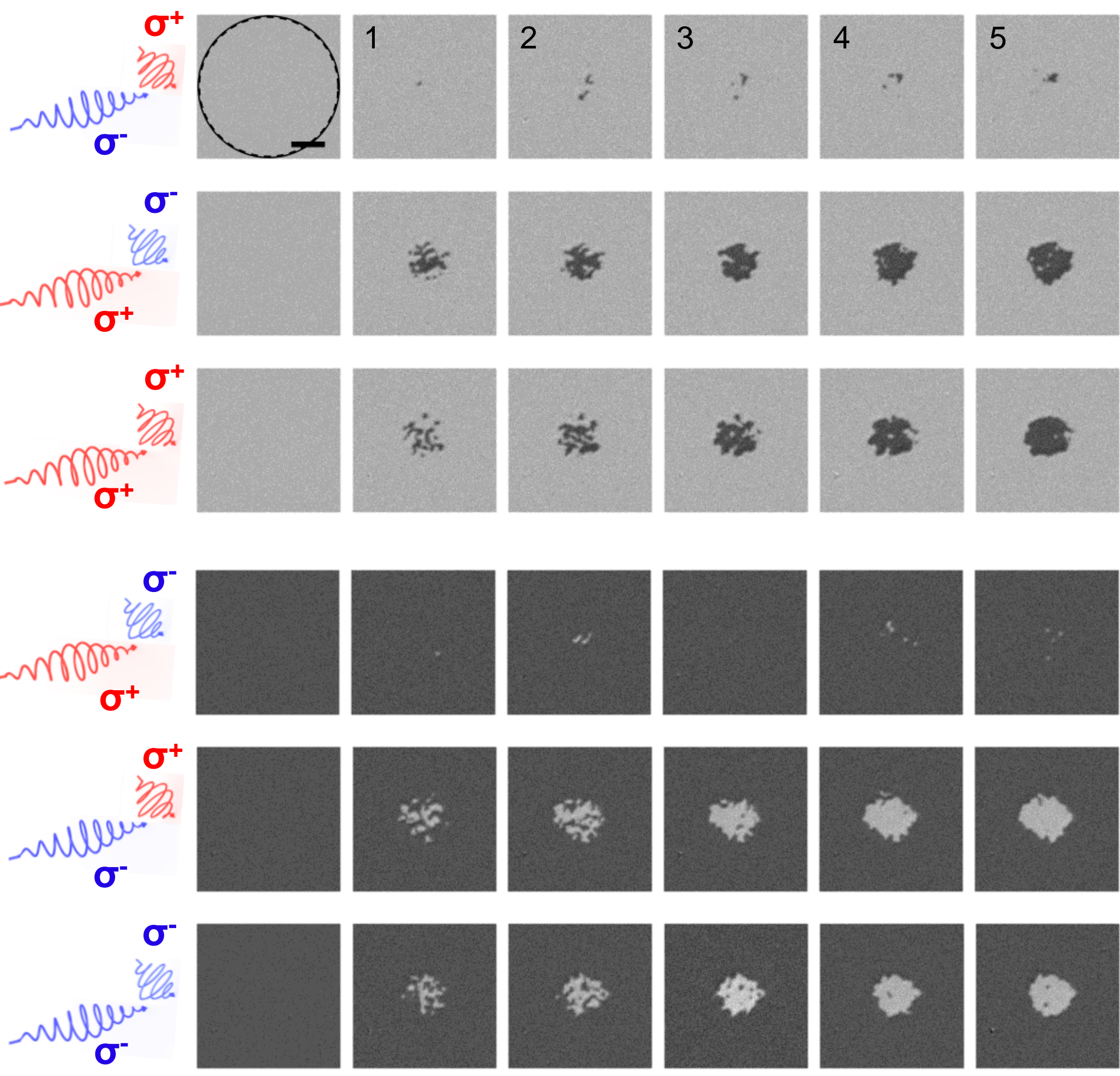

Figure 3 Yamada et al.